\documentclass[english]{elsart}
\usepackage[T1]{fontenc}
\usepackage[latin1]{inputenc}
\usepackage{amsmath}
\usepackage{graphicx}
\usepackage{amssymb}

\makeatletter

\providecommand{\tabularnewline}{\\}

\usepackage{graphicx}
\usepackage{psfig}

\usepackage{babel}
\makeatother
\begin{document}
\begin{frontmatter}

\title{Theory of two-photon absorption in poly(diphenyl) polyacetylenes }

\author{Alok Shukla}

\address{Physics Department, Indian Institute of Technology, Powai, Mumbai
400076 INDIA}

\ead{shukla@phy.iitb.ac.in}

\begin{abstract}
In this paper, we present a theoretical study of the nonlinear optical
response of the newly discovered conjugated polymer poly(diphenyl)polyacetylene
(PDPA). In particular, we compute the third-order nonlinear susceptibility
corresponding to two-photon absorption process in PDPA using: (a)
independent-particle H\"uckel model, and (b) using the correlated-electron
Pariser-Parr-Pople (P-P-P) model coupled with various configuration-interaction
methodologies such as the singles-configuration-interaction (SCI),
the multi-reference-singles-doubles CI (MRSDCI), and the quadruples-CI
(QCI) method. At all levels of theory, the polymer is found to exhibit
highly anisotropic nonlinear optical response, distributed over two
distinct energy scales. The low-energy response is predominantly polarized
in the conjugation direction, and can be explained in terms of chain-based
orbitals. The high-energy response of the polymer is found to be polarized
perpendicular to the conjugation direction, and can be explained in
terms of orbitals based on the side phenylene rings. Moreover, the
intensity of the nonlinear optical response is also enhanced as compared
to the corresponding polyenes, and can be understood in terms of reduced
optical gap.
\end{abstract}
\end{frontmatter}

\section{Introduction}

Conjugated polymers are among the prime candidates as materials for
the future nonlinear opto-electronic devices\cite{prasad}. The nonlinear
optical response of these materials owes its origin predominantly
to the $\pi$ electrons which (a) are localized as far as motion transverse
to the backbone of the polymer is concerned, but (b) are quite delocalized
along the backbone of the polymer. Indeed, the theoretical studies
of the nonlinear optical properties of a variety of conjugated polymers
such as \emph{trans}-polyacetylene, poly-(para)phenylene (PPP), poly-(para)phenylenevinylene
(PPV) etc. have attracted considerable attention over the past years\cite{agarwal,dixit,sumit,shakin,aparna,lavren,shukla-ppv}.
However, recently a new class of conjugated polymers called phenyl-substituted
polyacetylenes have been synthesized which are obtained by substituting
the side H atoms of \emph{trans}-polyacetylene by phenyl derivatives,
and their optical properties have been studied\cite{tada1,tada2,liess,fujii1,gontia,sun,hidayat,fujii2}.
The polymers obtained by replacing all the side H atoms of the \emph{trans}-polyacetylene
by phenyl groups are called phenyl-disubstituted polyacetylenes (PDPA's),
while the ones obtained by replacing alternate H atoms by the phenyl
groups are called poly-phenylacetylenes (PPA's). Experimentally it
was demonstrated that PDPA's exhibit strong photoluminescence (PL)
with large quantum efficiency\cite{liess,fujii1,gontia,fujii2}, while
the PPA's on the other hand exhibit weak PL\cite{hidayat}, akin to
\emph{trans}-polyacetylene. PDPA, similar to \emph{trans}-polyacetylene,
is a polymer with a degenerate ground state, therefore, strong PL
exhibited by it was considered to be counterintuitive\cite{gontia}.
However, in a series of papers we demonstrated that the strong PL
of PDPA's is due to reversed excited state ordering in these materials,
as compared to \emph{trans}-polyacetylene\cite{shukla1,shukla2,shukla3}.
In \emph{trans}-polyacetylene the two-photon state $2A_{g}$ occurs
below the optical state $1B_{u}$, rendering it a poor emitter, while
in PDPA's we showed that the reduced correlation effects stemming
from the delocalization of electrons along the transverse directions,
bring the $1B_{u}$ state below the $2A_{g}$, converting them into
strong emitters\cite{shukla1,shukla2,shukla3}. We also demonstrated
that due to the transverse delocalization, the optical gaps in PDPA's
get lowered as compared to \emph{trans}-polyacetylene\cite{shukla1,shukla2,shukla3}.
Additionally, we predicted that another consequence of transverse
delocalization will be the significant presence of the transverse
polarization ($y$-component, if the conjugation direction is $x$)
in the photon emitted during the PL process in PDPA's\cite{shukla1,shukla2,shukla3}.
Since then, this prediction of ours has been verified in oriented
thin-film based PL experiments conducted on PDPA, by Fujii et al\cite{fujii2}.

Despite numerous investigations of the linear optics of PDPA's, so
far there has neither been any theoretical, nor any experimental,
investigation of the nonlinear optical properties of these materials.
Intuitively it is obvious that, because of the large number of $\pi$-electrons
even in small oligomers of PDPA's, these systems should exhibit large
nonlinear optical response. Since PDPA's are materials which possess
inversion symmetry, therefore, similar to \emph{trans}-polyacetylene,
the first nonlinear response that they will exhibit will be at the
third order in the radiation field. However, unlike \emph{trans}-polyacetylene,
which has no side conjugation, the nonlinear response of PDPA's should
also be significant to the $y$-polarized radiation, thereby rendering
it anisotropic. Moreover, using linear spectroscopy, for centrosymmetric
systems such as PDPA's, it is possible to explore only the excited
states of $B_{u}$ symmetry. However, using two-photon absorption
(TPA) nonlinear spectroscopy, one can investigate $A_{g}$-type excited
states, thereby shedding more light on the nature of electronic states
in these systems. It is because of these reasons that we decided to
undertake a systematic theoretical study of the TPA spectroscopy of
PDPA oligomers. In the present paper, we consider oligo-PDPA's of
varying lengths and compute their third-order nonlinear susceptibilities
corresponding to the TPA process, using both the independent particle
H\"uckel model, as well as Coulomb-correlated Pariser-Parr-Pople
(P-P-P) mocel. We decided to perform both the independent-particle
as well as correlated electron calculations so as to compare and contrast
the predictions of both the models, as well as to understand the influence
of electron-correlation effects on the nonlinear optical response
of the PDPA's. For the P-P-P model calculations, we employed a configuration-interaction
(CI) methodology, and approaches such as singles-CI (SCI), multi-reference-singles-doubles
CI (MRSDCI), and quadruples-CI (QCI) were used to compute these spectra.
We also compare computed TPA spectra on oligo-PDPA's, with those on
the oligomers of \emph{trans}-polyacetylene (polyenes) with the same
number of unit cells. While performing this comparison, we pay particular
attention to the ``essential-states picture''\cite{dixit,sumit} which
was developed to explain the nonlinear optical response of conjugated
polymers such as \emph{trans}-polyacetylene, in terms of a small number
of excited states. Despite the fact that oligo-PDPA's are much more
complex in structure than polyenes, we find that the essential states
picture remains valid to a large extent for these systems as well.

The remainder of this paper is organized as follows. In section \ref{method}
we briefly describe the theoretical methodology used to perform the
calculations in the present work. Next in section \ref{results} we
present and discuss the calculated nonlinear optical susceptibilities
of oligo-PDPA's. Finally, in section \ref{conclusion} we summarize
our conclusions, and discuss possible directions for future work.

\section{Methodology}

\label{method}

\begin{figure}
\begin{center}\includegraphics[%
  width=4cm]{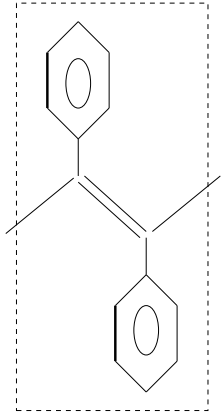}\end{center}

\caption{The unit cell of PDPA. The phenyl rings are rotated with respect
to the $y$-axis, which is transverse to the axis of the polyene backbone
($x$-axis)}

\label{fig-pdpa}
\end{figure}

The unit cell of PDPA oligomers considered in this work is presented
in Fig. \ref{fig-pdpa}%
 Ground state geometry of PDPA's, to the best of our knowledge, is
still unknown. However, from a chemical point of view, it is intuitively
clear that the steric hindrance would cause a rotation of the side
phenyl rings so that they would no longer be coplanar with the polyene
backbone of the polymer. The extent of this rotation is also unknown,
however, it is clear that the angle of rotation has to be less than
90 degrees because that would effectively make the corresponding hopping
element zero, implying a virtual disconnection of the side phenyl
rings from the backbone. In our previous works\cite{shukla1,shukla2,shukla3},
we argued that the steric hindrance effects can be taken into account
by assuming that the phenyl rings of the unit cell are rotated with
respect to the $y$-axis by 30 degrees in such a manner that the oligomers
still have inversion symmetry. In the following, we will adopt the
notation PDPA-$n$ to denote a PDPA oligomer containing $n$ unit
cells of the type depicted in Fig. \ref{fig-pdpa}.

The point group symmetry associated with \emph{trans}-polyacetylene
(and polyenes) is $C_{2h}$ so that the one-photon states belong to
the irreducible representation (irrep) $B_{u}$, while the ground
state and the two-photon excited states belong to the irrep $A_{g}$.
Because of the phenyl group rotation mentioned above, the point group
symmetry of PDPA's is $C_{i}$ so that its ground state and the two-photon
excited states belong to the irrep $A_{g}$, while the one-photon
excited states belong to the irrep $A_{u}$. However, to facilitate
direct comparison with polyenes, we will refer to the one-/two-photon
states of PDPA's also as $B_{u}$/$A_{g}$-type states.

The independent-electron calculations on the oligomers PDPA-$n$ were
performed using the H\"{u}ckel model Hamiltonian which, adopting
a notation identical to our previous works\cite{shukla1,shukla2,shukla3}
reads, \begin{subequations}
\label{allequations}
\begin{equation}
H = H_C + H_P + H_{CP},  \label{eq-ham}
\end{equation}
where \(H_C\) and \(H_P\) are the one-electron Hamiltonians for the carbon atoms 
located on the {\emph{trans}}-polyacetylene backbone (chain), and the  phenyl groups, respectively, 
H\(_{CP}\) is the one-electron hopping between the chain and the phenyl
units. The individual terms can now be written as, 
\begin{equation}
H_C = -\sum_{\langle k,k' \rangle,M} (t_0 - (-1)^M \Delta t) 
B_{k,k';M,M+1},
  \label{eq-h1}
\end{equation}
\begin{equation}
H_P=-t_0 \sum_{\langle\mu,\nu\rangle,M} B_{\mu,\nu;M,M}, \label{eq-h2}
\end{equation}
and 

\begin{equation}
H_{CP}= -t_{\perp} \sum_{\langle k,\mu \rangle,M} B_{k,\mu;M,M,}. \label{eq-h3}
\end{equation}
\end{subequations} In the equation above, $k$, $k'$ are carbon atoms on the polyene
backbone, $\mu,\nu$ are carbon atoms located on the phenyl groups,
$M$ is a unit consisting of a phenyl group and a polyene carbon,
$\langle...\rangle$ implies nearest neighbors, and $B_{i,j;M,M'}=\sum_{\sigma}(c_{i,M,\sigma}^{\dagger}c_{j,M',\sigma}+h.c.)$.
Matrix elements $t_{0}$, and $t_{\perp}$ depict one-electron hops.
In $H_{C}$, $\Delta t$ is the bond alternation parameter arising
due to electron-phonon coupling. In $H_{CP}$, the sum over $\mu$
is restricted to atoms of the phenyl groups that are directly bonded
to backbone carbon atoms.

As far as the values of the hopping matrix elements are concerned,
we took $t_{0}=2.4$ eV, while it is imperative to take a smaller
value for $t_{\perp}$, because of the twist in the corresponding
bond owing to the steric hindrance mentioned above. We concluded that
for a phenyl group rotation of 30 degrees, the maximum possible value
of $t_{\perp}$ can be 1.4 eV\cite{shukla1}. Bond alternation parameter
$\Delta t=0.45$ eV was chosen so that the backbone corresponds to
\emph{trans}-polyacetylene with the optical gap of 1.8 eV in the long
chain limit.

The TPA processes in oligo-PDPA's were studied by computing the third-order
nonlinear susceptibilities $\chi^{(3)}(-\omega;\omega,-\omega,\omega)$.
For short we will refer to these susceptibilities as $\chi_{TPA}^{(3)}$.
First the H\"{u}ckel Hamiltonian for the corresponding oligo-PDPA
was diagonalized to compute the one-electron eigenvalues and eigenfunctions.
These quantities were subsequently used in the formulas derived by
Yu and Su\cite{yu-su}, to compute $\chi_{TPA}^{(3)}$. 

The correlated calculations on oligo-PDPA's were performed using the
P-P-P model Hamiltonian \begin{equation}
H=H_{C}+H_{P}+H_{CP}+H_{ee},\label{eq-ham2}\end{equation}
 where $H_{C}$, $H_{P}$, $H_{CP}$ are the one-electron parts of
the Hamiltonian mentioned above, while H$_{ee}$ depicts the electron-electron
repulsion and can be written as \begin{eqnarray}
H_{ee} & = & U\sum_{i}n_{i\uparrow}n_{i\downarrow}\nonumber \\
 &  & +\frac{1}{2}\sum_{i\neq j}V_{i,j}(n_{i}-1)(n_{j}-1),\label{eq-hee}\end{eqnarray}
 where $i$ and $j$ represent all the atoms of the oligomer. The
Coulomb interactions are parameterized according to the Ohno relationship
\cite{ohno}, \begin{equation}
V_{i,j}=U/\kappa_{i,j}(1+0.6117R_{i,j}^{2})^{1/2}\;\mbox{,}\label{eq-ohno}\end{equation}

where, $\kappa_{i,j,M,N}$ depicts the dielectric constant of the
system which can simulate the effects of screening, $U$ is the on-site
repulsion term, and $R_{i,j}$ is the distance in \AA ~ between
the $i$th carbon and the the $j$th carbon. The values of $U$ and
$\kappa_{i,j,M,N}$ are unknown quantities and, at present, it is
not clear as to what range of values of these parameters are suitable
for PDPA's. In our earliear studies of linear optics of these materials
we tried two parameter sets: (a) {}``standard parameters'' with
$U=11.13$ eV and $\kappa_{i,j}=1.0$, and (b) {}``screened parameters''
with $U=8.0$ eV and $\kappa_{i,i,M,M}=1.0$, and $\kappa_{i,j,M,N}=2.0$,
otherwise\cite{shukla1,shukla2,shukla3}. Using the screened parameters,
Chandross and Mazumdar\cite{chandross} obtained better agreement
with experiments on excitation energies of PPV oligomers, as compared
to the standard parameters. Recently, we performed a large-scale correlated
study of singlet and triplet excited states in oligo-PPV's and observed
a similar trend\cite{shukla-ppv}. As far as the hopping matrix elements
are concerned, for correlated calculations we used the same values
for these parameters as in the independent electron calculations except
for the value of $\Delta t$ which was taken to be 0.168 eV. 

The starting point of the correlated calculations for various oligomers
were the restricted Hartree-Fock (HF) calculations, using the P-P-P
Hamiltonian. The many-body effects beyond HF were computed using different
levels of the configuration interaction (CI) method, namely, singles-CI
(SCI), quadruples-CI (QCI), and the multi-reference singles-doubles
CI (MRSDCI). Since the number of electrons in oligo-PDPA's is quite
large despite the P-P-P approximation owing to the large unit cell
(fourteen electrons/cell), except for the SCI calculations, it is
not possible to include all the orbitals in the many-body calculations.
Therefore, one has to reduce the number of degrees of freedom by removing
some orbitals from the many-body calculations. In order to achieve
that, for each oligomer we first decided as to which occupied and
the virtual orbitals will be active in the many-body calculations
based upon: (a) their single-particle HF energies with respect to
the location of the Fermi level, and (b) Mulliken populations of various
orbitals with respect to the chain/phenylene-based atoms. Because
of the particle-hole symmetry in the problem, the numbers of active
occupied and virtual orbitals were taken to be identical to each other,
with the occupied and virtual orbitals being particle-hole symmetric.
The remaining occupied orbitals were removed from the many-body calculations
by the act of {}``freezing'', i.e., by summing up their interactions
with the active electrons, and adding this effective potential to
the one-electron part of the total Hamiltonian. The inactive virtual
orbitals were simply deleted from the list of orbitals. When we present
the CI results on various oligo-PDPA's, we will also identify the
list of active orbitals. During the CI calculations, full use of the
spin and the point group ($C_{i}$ for PDPA's) symmetries was made.
From the CI calculations, we obtain the eigenfunctions and eigenvalues
corresponding to the correlated ground and excited states of various
oligomers. Using the many-body wave functions, we compute the matrix
elements of the dipole operator amongst various states. Finally, these
quantities are fed in to the sum-over-states formulas of Orr and Ward,~\cite{orr}
to obtain the correlated values of the TPA susceptibility $\chi^{(3)}(-\omega;\omega,-\omega,\omega)$.
More details about the procedural aspects of various CI approaches
used by us can be found in our earlier works\cite{shukla-ppv,shukla2,shukla3,hng-ppv}.

\section{Results and Discussion}

\label{results} In this section we present our results on the TPA
susceptibilities of oligo-PDPA's, computed using various approaches.
Where applicable, we also compare our results for the oligo-PDPA's
to the corresponding results on polyenes of the same conjugation length.

\subsection{Independent-Electron Theory}

In this section we briefly present and discuss the TPA spectra of
oligo-PDPA's computed using the independent-particle H\"uckel model.
First we present the results on the longitudinal component of the
susceptibility, followed by those on the transverse component.

\subsubsection{Longitudinal Component}

\label{res-tpa-x}

\begin{figure}
\begin{center}\includegraphics[%
  width=8cm,
  angle=-90]{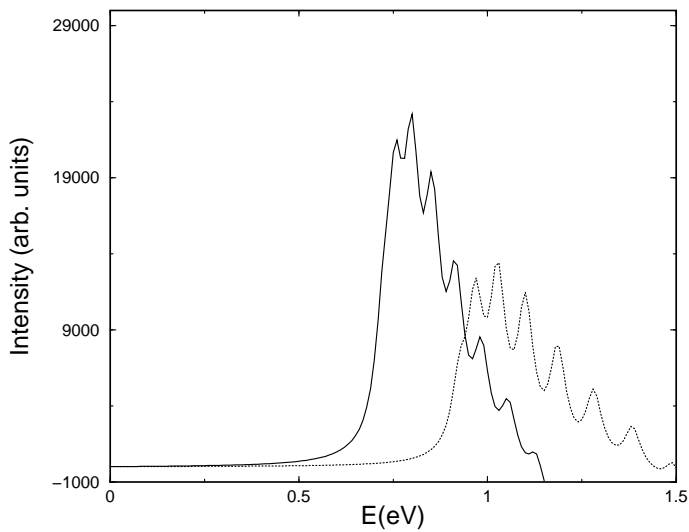}\end{center}

\caption{Comparison of the imaginary part of $\chi_{xxxx}^{(3)}(-\omega;\omega,-\omega,\omega)$
of PDPA (solid line) and \emph{trans}-polyacetylene (dotted line)
oligomers containing fifty unit cells, computed using the H\"{u}ckel
model. With our choice of hopping parameters (see text) optical gap
of PDPA-50 was computed to be 1.42 eV, while that of the \emph{trans}-polyacetylene
oligomer was 1.82 eV. A linewidth of 0.05 eV was assumed for all energy
levels.}

\label{tpa-pdpa-tpa}
\end{figure}

\begin{figure}
\begin{center}\includegraphics{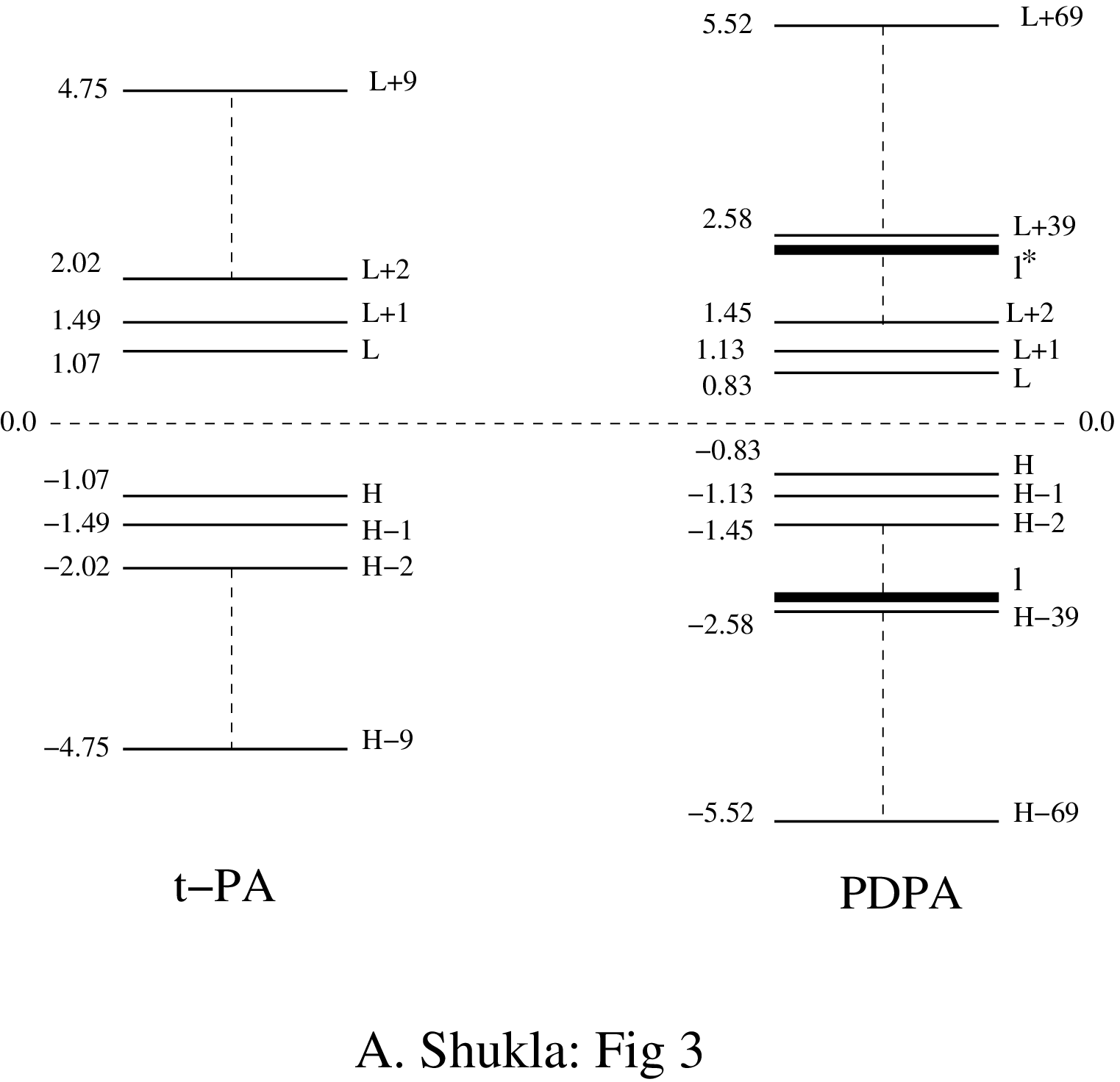}\end{center}

\caption{Important single particle energy levels of ten unit oligomers of
\emph{trans}-polyacetylene (labeled t-PA) and PDPA computed using
the H\"uckel model. Energy levels are drawn to the scale, with their
values (in eV) indicated next to them. Chemical potential is indicated
as the dotted line through the center, and is assumed to be  0.0 eV.
For the chosen hopping parameters (see text), the optical gap in PDPA
oligomer is 1.66 eV as compared to 2.14 eV in the t-PA oligomer. As
per the notation explained in the text, H/L indicate HOMO/LUMO orbitals.
The thick lines labaled $l/l^{*}$denote the localized orbitals of
phenyl rings, all of which are located at -2.4/2.4 eV. In addition
to the optical gap, narrowing of other energy gaps in PDPA, as compared
to t-PA, is obvious.\label{fig-levels}}
\end{figure}

The longitudinal component of the two-photon absorption spectrum,
$\chi_{xxxx}^{(3)}(-\omega;\omega,-\omega,\omega)$, for oligo-PDPA's
describes the longitudinal nonlinear optical response of the material
to the radiation polarized along the conjugation direction ($x$-axis).~\cite{chi-explan}
It is represented by the imaginary part of $\chi_{xxxx}^{(3)}(-\omega;\omega,-\omega,\omega)$,
calculated values of which for PDPA-50 are presented in Fig.\ref{tpa-pdpa-tpa}.
For the sake of comparison, the same figure also presents the longitudinal
TPA spectrum of fifty unit cell polyene. Although the plots of the
TPA spectra of the two materials appear qualitatively similar, however,
there are significant quantitative differences between the two. It
is obvious from Fig. \ref{tpa-pdpa-tpa} that: (a) The magnitude of
the resonant nonlinear response in PDPA is significantly enhanced
as compared to the \emph{trans}-polyacetylene, and (b) all the resonant
features of the two-photon spectrum of the PDPA are significantly
redshifted as compared to \emph{trans}-polyacetylene. Although, in
Fig.\ref{tpa-pdpa-tpa} the TPA spectra of relatively larger oligomers
of the two substances are compared, the same trend is obvious even
for small oligomer such as PDPA-5, when compared to a polyene of the
same size. Thus H\"{u}ckel model calculations on both small and large
oligomers suggest that the resonant parts of the longitudinal component
of the TPA of PDPA's are significantly more intense, and redshifted,
compared to \emph{trans}-polyacetylene. This point can be further
understood by referring to Fig. \ref{fig-levels}, in which the energy
levels of a ten unit polyene and PDPA-10 are presented side by side.
From the figure it is obvious that all comparable energy gaps, including
the optical gap, are narrowed in PDPA, as compared to the corresponding
polyene.

Next we study the many-particle states which are visible in the TPA
spectrum of various oligomers of PDPA. For the purpose, we denote
the highest occupied molecular orbital (HOMO) as $H$, and the lowest
unoccupied molecular orbital (LUMO) as $L$. For PDPA-10, the most
intense peak corresponds to the $2A_{g}$ state of the oligomer obtained
by $|H\rightarrow L+1\rangle$ and $|H-1\rightarrow L\rangle$ singlet
excitations (see Fig. \ref{fig-levels} for the energy levels). However,
as the size of the oligomer increases, the $2A_{g}$ state ceases
to be the most intense peak. For example, in the TPA spectrum of PDPA-30,
the $3A_{g}$ state corresponding to the $|H\rightarrow L+3\rangle$
and $|H-3\rightarrow L\rangle$ excitations is the most intense peak.
This trend continues with the increasing size of the oligomer, and
we speculate that the longitudinal TPA spectrum of an infinite PDPA
would be a broad resonance, with several $A_{g}$-type states contributing
to the intensity, in perfect qualitative agreement with the TPA spectrum
of infinite \emph{trans}-polyacetylene. \cite{agarwal} The other
noticeable aspect of these $A_{g}$-type states is that they all originate
from the excitations among the orbitals close to the Fermi level,
which have polyene-like character, with large Mulliken population
of sites located on the backbone. Thus, we conclude that, as far as
the independent-electron theory is concerned, the longitudinal TPA
spectrum of PDPA is qualitatively similar to that of \emph{trans}-polyecetylene.


\subsubsection{Transverse Component}

\label{res-tpa-y} 
\begin{figure}
\begin{center}\includegraphics[%
  width=8cm,
  angle=-90]{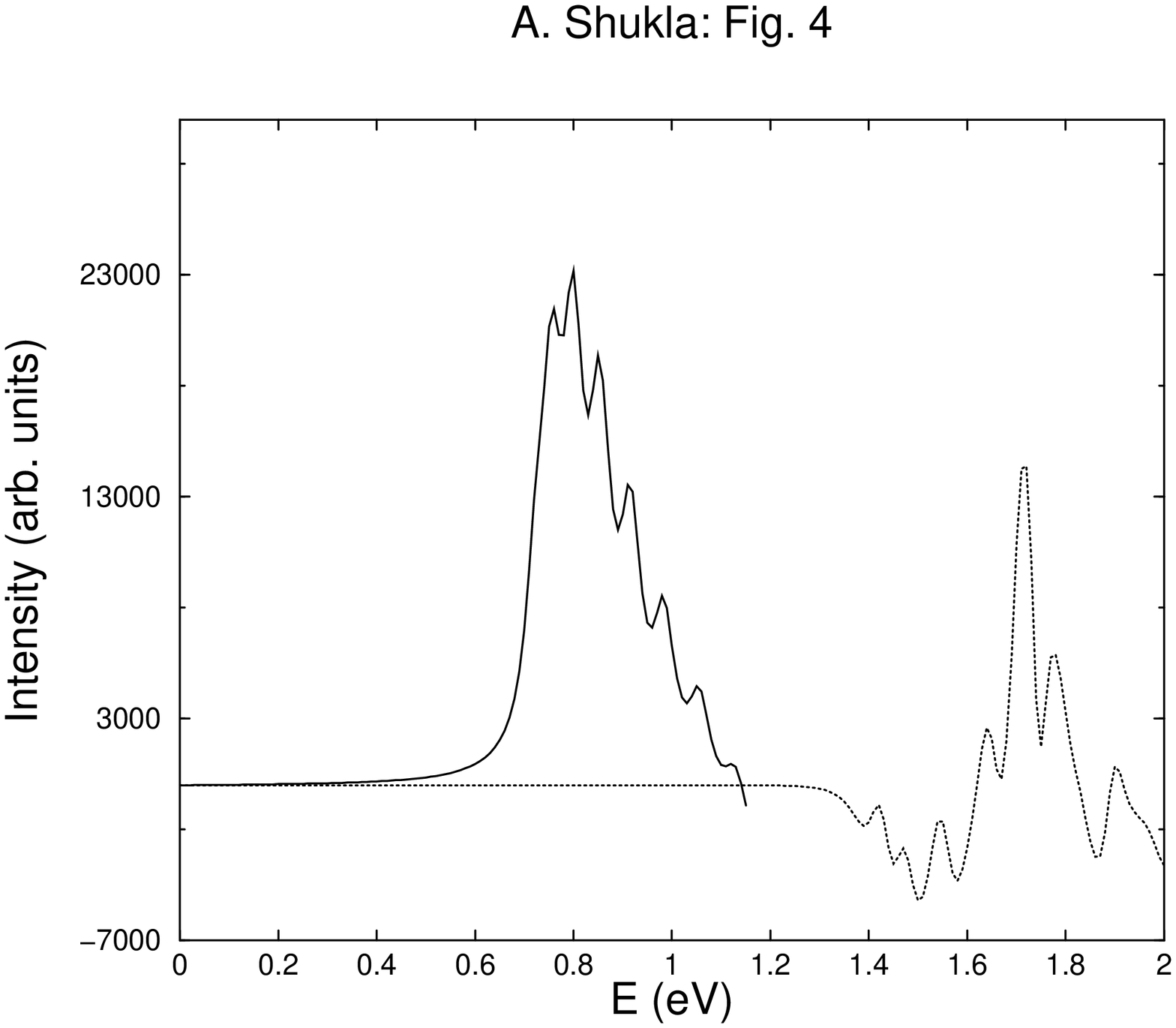}\end{center}

\caption{Comparison of the longitudinal (solid lines) and the transverse (dotted
lines) components of the TPA susceptibilities for a PDPA oligomer
containing fifty unit cells, computed using the H\"{u}ckel model.
Linewidth of 0.05 eV was assumed for all energy levels.}

\label{tpa-pdpa-x-y}
\end{figure}

Imaginary part of the susceptibility component
$\chi_{yyyy}^{(3)}(-\omega;\omega,-\omega,\omega)$ describes the
lowest-order nonlinear optical absorption of the transversely polarized
radiation in oligo-PDPA's. It is this component of the susceptibility
which distinguishes the nonlinear response of PDPA from that of \emph{trans}-polyacetylene.
Clearly, due to minimal extension in the transverse direction, \emph{trans}-polyacetylene
will have negligible values of this component. In Fig. \ref{tpa-pdpa-x-y}
we plot on the same scale, both the longitudinal, and the transverse,
components of the TPA susceptibility for PDPA-50 computed using the
H\"uckel model.%

From Fig. \ref{tpa-pdpa-x-y}, it is clear that: (a) the resonant
intensities of the transverse component of the susceptibility within
the H\"{u}ckel model are comparable to that of the longitudinal ones,
(b) the excited states contributing to the transverse component are
much higher in energy as compared to the ones contributing to the
longitudinal component. Moreover, the transverse component displays
very weak size dependence as far as the location of the main peak
is concerned which, for all the oligomers studied, is roughly close
to $1.7$ eV. The weak size dependence of the resonant energy suggests
that the excited states in question must involve phenyl derived single-electron
levels, because with the increasing conjugation length of the oligomer,
its size in the transverse direction remains unchanged. Next we investigate
the energy levels involved in the transverse TPA susceptibilities
of PDPA's.

Since the number of one-electron levels proliferates tremendously
with the increasing conjugation length due to band formation, it is
fruitful first to investigate smaller oligomers such as PDPA-10. In
PDPA-10, the main peak occurs at 1.71 eV which corresponds to an $A_{g}$
type state obtained from the ground state by single-excitations $H\rightarrow L+39$
and $H-39\rightarrow L$. To investigate the nature of the orbital
corresponding to the $L+39$-th one-electron state, we calculated
the contribution of the charge density centered on the backbone carbon
atoms, to its total normalization. The contribution was computed to
be $0.05$ which indicates that the orbital in question is predominantly
centered on the side phenyl rings. Further investigation of the orbital
coefficient reveals that the orbitals in question, are derived from
the phenyl-based delocalized ($d^{*})$ virtual orbitals, with no
contribution from the localized orbitals ($l^{*}$-type) of the phenyl
rings. Similarly, owing to the particle-hole symmetry, $H-39$-th
orbital is derived from the $d$-type occupied orbitals of the phenyl
rings. The energy levels corresponding to these orbitals are presented
in Fig. \ref{fig-levels}.

Next we examine the case of PDPA-50, for which the highest peak is
located at 1.72 eV. Here we find that most of the states in this energy
region correspond to single-excitations $H-5\rightarrow L+153+i$
(and its particle-hole reversed counterpart) with $\{ i=0,\ldots,8\}$.
Again a charge density analysis of these levels reveals that while
the levels $H-5$ (and $L+5$) have chain atom contribution of $\approx0.72$,
while the levels $L+153+i$ (and $H-153-i$) have chain atom contribution
to the total normalization $\approx0.01$. Thus it is clear that $H-5$
($L+5$) orbital constitutes top of the valence band (bottom of the
conduction band) composed of the chain levels, while the orbital coefficients
reveal that the levels $H-153-i$ ($L+153+i$) are part of the occupied
(unoccupied) bands composed predominantly of the phenyl-based $d$
($d^{*}$) levels. Thus it is clear that the transverse TPA susceptibility
of the system owes its origins to phenyl-based levels. This is an
important point which also helps us perform effective correlated calculations
of this component, presented next.

\subsection{Correlated-Electron Theory}

\label{sub-corr}

Based upon correlated-electron calculations performed on polyenes,
a simplified picture of their nonlinear optical properties---referred
to us ``essential-states picture''---has emerged\cite{dixit,sumit}.
According to this picture, most of the nonlinear optical properties
of polyenes can be understood based upon the excited states $1B_{u}$
(the lowest one-photon state), $mA_{g}$ (a state with strong dipole
coupling to $1B_{u}$), and $nB_{u}$ (a state with strong dipole
coupling to $mA_{g}$). Although, the state $2A_{g}$ (a two-photon
state lower in energy than $1B_{u}$ in polyenes) also has a large
dipole coupling to the $1B_{u}$ state, its contribution to nonlinear
susceptibilities becomes very small due to some curious cancellation
effects resulting from the many-body nature of these states\cite{dixit,sumit}.
For the particular case of TPA susceptibility, in polyenes one observes
a weak peak corresponding to the $2A_{g}$ state, while the most intense
peak is derived from the $mA_{g}$ state arising from the excitation
channel $1A_{g}\rightarrow1B_{u}\rightarrow mA_{g}\rightarrow1B_{u}\rightarrow1A_{g}$\cite{dixit,sumit}.
Because of the structural similaritiy of PDPA to \emph{trans}-polyacetylene,
while presenting our results we will analyze them from the standpoint
of the essential states mechanism. Moreover, since the transverse
nonlinear spectrum of these materials is completely novel as compared
to the polyenes, one wonders whether an essential states mechanism
also holds for this component as well.

First, however, we briefly elucidate the role played by the choice
of Coulomb parameters in the P-P-P model, because, as mentioned earlier
correct parameters are still unknown for PDPA's, thereby making the
choice of these parameters very important.

\subsubsection{Role of Coulomb parameters}

\label{sub-param}

\begin{figure}
\begin{center}\includegraphics[%
  width=8cm,
  angle=-90]{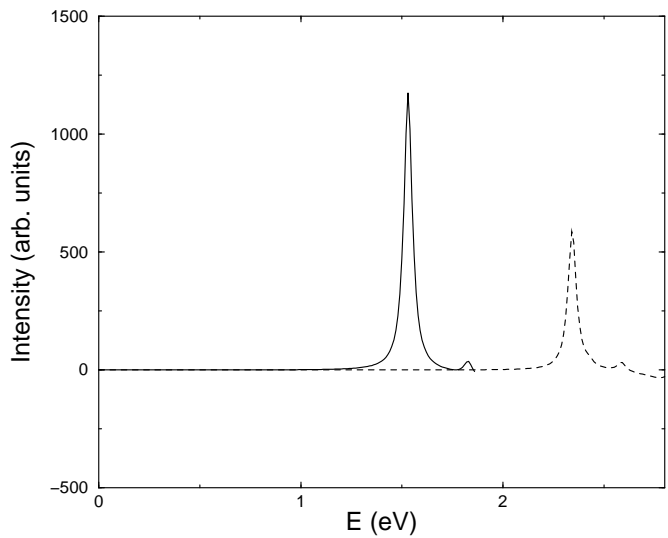}\end{center}

\caption{Comparison of imaginary parts of $\chi_{xxxx}^{(3)}(-\omega;\omega,-\omega,\omega)$
spectra of PDPA-5 computed with screened parameters (solid line) and
standard parameters (broken line). A linewidth of 0.05 eV was assumed
and the SCI method was used to compute the excited states.}

\label{fig-par}
\end{figure}

In the present work, we have performed the calculations using both
the standard Ohno parameters, as well as the screened parameters of
Chandross and Mazumdar\cite{chandross} to describe the P-P-P Hamiltonian.
However, based upon our earlier calculations dealing with the optical
properties of PDPA\cite{shukla1,shukla2,shukla3}, and PPV\cite{shukla-ppv,hng-ppv},
the screened-parameter-based calculations generally yield much better
agreement with experiments than the ones based upon the standard parameters.
The standard-parameter-based calculations for such systems generally
yield energy gaps which are larger than the experimental ones. In
Fig.\ref{fig-par} we compare the longitudinal components of the TPA
spectra ($\chi_{xxxx}^{(3)}(-\omega;\omega,-\omega,\omega)$) for
PDPA-5 computed both with the standard and the screened parameters,
using the SCI approach.%
 It is clear from the figure that qualitatively speaking both the
spectra are similar. However, from a quantitative viewpoint: (a) the
intensities are significantly larger as computed with the screened
parameters, (b) and more importantly the resonant features of the
screened spectrum are substantially red-shifted as compared to the
standard spectrum.

Since the smaller energy gaps obtained with the screened parameters
in our earlier works were found to be in much better agreement with
the experiments, we will present our main results based upon screened-parameter-based
calculations. However, when we compare the PDPA nonlinear optical
spectra with those of polyenes, we will use the standard parameters
because screened parameters are not valid for polyenes.

\label{res-tpa}

\subsubsection{Longitudinal Component}

\label{res-tpa-corr-x}

\begin{figure}
\begin{center}\includegraphics[%
  width=12cm,
  angle=-90]{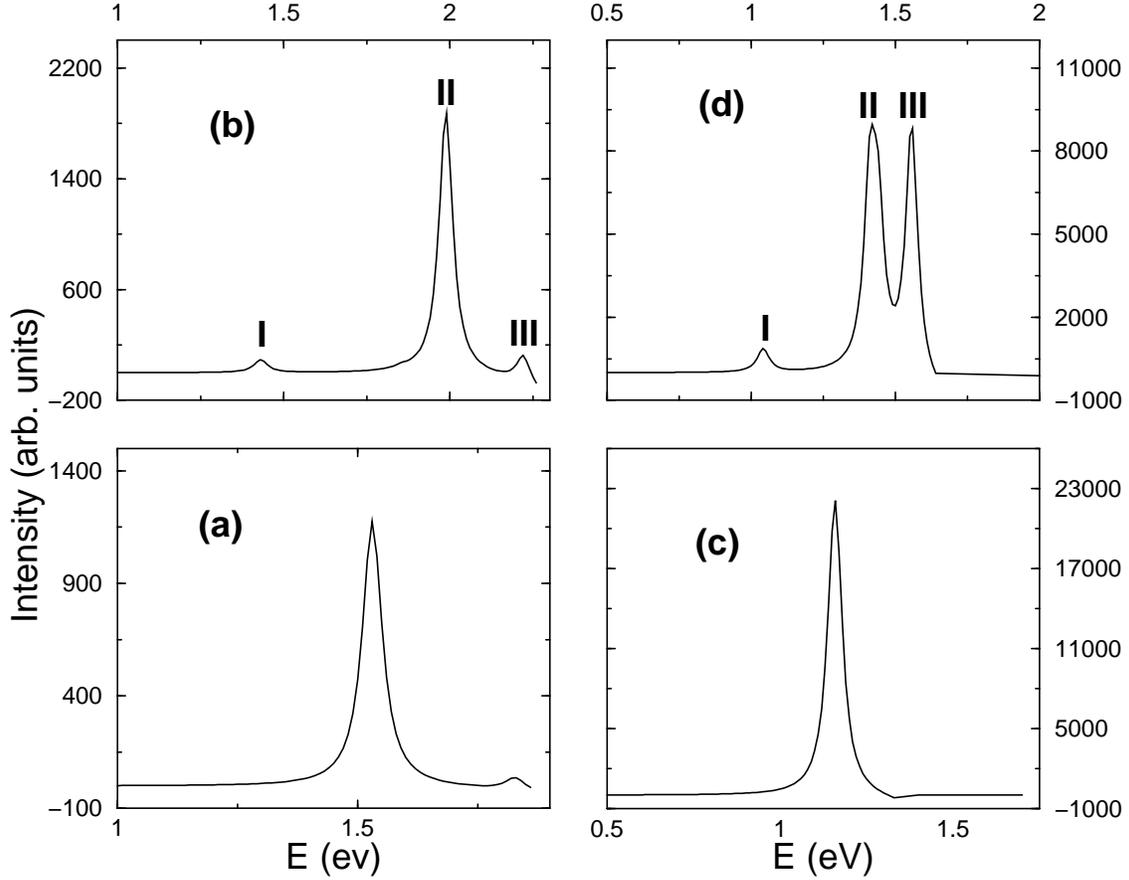}\end{center}

\caption{Imaginary parts of the longitudinal two-photon absorption spectra
($\chi_{xxxx}^{(3)}(-\omega;\omega,-\omega,\omega)$) of oligo-PDPA's
calculated using various CI approaches and the screened parameters:
(a) PDPA-5 (SCI), (b) PDPA-5 (QCI), (c) PDPA-10 (SCI), and (d) PDPA-10
(QCI). A common linewidth of 0.05 eV was assumed for all the levels.}

\label{pdpa-all-tpa}
\end{figure}

Imaginary parts of $\chi_{xxxx}^{(3)}(-\omega;\omega,-\omega,\omega)$,
calculated with screened parameters and the SCI and QCI approaches
for PDPA-5 and PDPA-10 are presented in Fig.\ref{pdpa-all-tpa}.

Energies and many-particle wave-functions of various $A_{g}$-type
excited states contributing to peaks in different spectra are summarized
in table \ref{tab-pdpa5} for PDPA-5 and in table \ref{tab-pdpa10}
for PDPA-10.

 First, we will discuss the results obtained using the SCI approach,
presented in Figs. \ref{pdpa-all-tpa}(a), and \ref{pdpa-all-tpa}(c).
Although, the SCI approach does not include the electron correlation
effects to a high order, it does include them in a more balanced way
for the ground and the excited states as compared to the singles-doubles-CI
(SDCI) approach. Therefore, it provides a feel for the influence of
electron correlation effects on various properties. Moreover, since
it is computationally not very expensive, it allows us to include
all the orbitals in the calculations which is certainly not possible
for large systems such as oligo-PDPA's if some more extensive approach
such as MRSDCI/QCI were used. An inspection of the spectra reveals
the following general features: (i) The susceptibilities of both the
oligomers have only one intense peak while, with the increasing sizes
of oligomers, as expected, the peak is redshifted and becomes much
more intense.

Next we examine the SCI many-particle wave functions of the states
contributing to the peaks in the longitudinal TPA spectra. For PDPA-5
(Fig. \ref{pdpa-all-tpa}(a)), the peak in the TPA spectrum corresponds
to the $2A_{g}$ state located at 3.06 eV whose SCI wave function
gets its main contributions from the singly excited configurations
$|H\rightarrow L+1\rangle$ and $|H-1\rightarrow L\rangle$ with coefficients
0.69 denoted in short as $|H\rightarrow L+1\rangle+c.c.$ (0.69) in
table \ref{tab-pdpa5}\cite{notation}. Many other singly-excited
configurations make smaller contributions to the total wave function.
For the case of PDPA-10 (Fig. \ref{pdpa-all-tpa}(c)), as is clear
from table \ref{tab-pdpa10}, the peak again corresponds to the $2A_{g}$
state, located at a lower energy, has a many-particle wave function
qualtitatively identical to that for PDPA-5. Since at the SCI level,
the many-particle wave function is missing two-particle excitations,
therefore, the $2A_{g}$ state actually corresponds to the $mA_{g}$
state obtained at a higher energy in more sophisticated calculations
such as the QCI\cite{sum-priv}. Clearly, the picture emerging from
the SCI calculations is that it is the $A_{g}$-type excited states
resulting from single-particle excitations among the orbitals close
to the Fermi level which contribute to the main intensity in the longitudinal
TPA spectrum. Clearly, this picture is in very good agreement with
the H\"{u}ckel-model-based results reported in the previous section.

Therefore, now it is of interest to explore as to how high-order correlation
treatments such as the QCI approach influence these properties. However,
given the large number of electrons in these systems, QCI method is
not feasible for them if all the orbitals of the system are retained
in the calculations. Since the longitudinal nonlinear optical properties
are determined by low-lying excited states of the system, in the limited
CI calculations we decided to include the orbitals closest to the
Fermi level. Therefore, for PDPA-$n$ we included $n$ occupied, and
$n$ virtual orbitals closest to the Fermi level in the QCI calculations.
Remaining occupied orbitals were frozen and virtual orbitals were
deleted as explained in section \ref{method}. Thus, the computational
effort associated with the QCI calculations on PDPA-$n$ is same as
that needed for a polyene with $n$ double bonds. Although for PDPA-$10$,
it leads to Hilbert space dimensions in excess of one million, however,
using the methodology reported in our earlier works\cite{shukla-ppv,shukla2},
we were able to obtain low-lying excited states of such systems.

In Figs. \ref{pdpa-all-tpa}(b) and \ref{pdpa-all-tpa}(d) we present
the longitudinal TPA susceptibilities of PDPA-5 and PDPA-10 computed
using the QCI approach, and utilizing the screened-parameters in the
P-P-P Hamiltonian. It is clear from the figures that that the longitudinal
TPA spectra at the QCI/screened level exhibit three peaks (labeled
I, II, and III) while the corresponding SCI spectrum presented earlier
exhibited only one peak. Peak I is a weak feature which corresponds
to the $2A_{g}$ states of both the oligomers. Due to the better treatment
of electron correlation effects, one obtains a $2A_{g}$ state which
is distinct from the $mA_{g}$ state both in excitation energy, and
contribution to the TPA spectrum. The $2A_{g}$ state, in complete
analogy with the situation in polyenes, is mainly composed of low-lying
single and double excitations and makes a negligible contribution
to the longitudinal TPA spectra of oligo-PDPA's. This aspect of electron
correlation effects has been observed over the years in the calculations
on polyenes\cite{sumit}, and recently on PPP and PPV polymers as
well\cite{hng-ppv}, where the contribution of the $2A_{g}$ state
to the TPA spectrum diminishes once sophisticated many-body techniques
are employed to describe the excited states.

Feature II of PDPA-5 is due to the $4A_{g}$ state of the oligomer,
while in PDPA-10 it derives its intensity from two nearly degenerate
states $3A_{g}$ and $4A_{g}$. Since this feature corresponds to
the most intense peak in the spectrum, based on that that crieterion
alone, one is tempted to label it as the $mA_{g}$ state of the corresponding
oligomer. When we compare the many-particle state of $mA_{g}$ to
that of $2A_{g}$, we observe that the same configurations ($|H\rightarrow L;H\rightarrow L\rangle$
and $|H-1\rightarrow L+2\rangle+c.c$) make the most important contributions
to both the states, a feature again in common with the trend seen
in polyenes\cite{sumit}, and PPP and PPV\cite{hng-ppv}. Finally,
when we examine the longitudinal dipole couplings of these states
with the $1B_{u}$ state, we observe that it is much higher for the
$mA_{g}-1B_{u}$ coupling as compared to the $2A_{g}-1B_{u}$ coupling,
clearly explaining the relative intensity pattern of the longitudinal
TPA spectra of oligo-PDPA's. Based upon this we conclude that the
feature II of the spectrum indeed corresponds to the $mA_{g}$ states
of the respective oligomers.

Finally, feature III in PDPA-5 is a rather weak feature while, in
PDPA-10, it is a very intense peak. In both the oligomers this feature
is due to the $5A_{g}$ states whose many-particle wave function is
listed in tables \ref{tab-pdpa5} and \ref{tab-pdpa10} are qualitatively
similar. In both the oligomers this state is a linear combination
of higher-energy singly- and doubly-excited configurations. However,
the fact that, contrary to PDPA-5, in PDPA-10 this state has a strong
dipole coupling to the $1B_{u}$ state clearly points to the band
formation and suggests the possibility that in longer oligomers this
feature will merge with feature II leading to a single $mA_{g}$ band. 

\begin{table}

\caption{Nature of $A_{g}$ type states contributing to the longitudinal TPA
spectrum of PDPA-5 as obtained in various CI calculations. Under the
heading wave function, we list the most important configurations contributing
to the many-body wave function of the state concerned, along with
their coefficients, consitent with our convention\cite{notation}. }

\begin{tabular}{|c|c|c|c|c|}
\hline 
Calculation&
Feature&
State&
Energy (eV)&
Wave Function\tabularnewline
\hline
\hline 
SCI&
I&
$2A_{g}$($mA_{g}$)&
3.06&
$|H\rightarrow L+1\rangle+c.c\:(0.69)$\tabularnewline
\hline 
QCI&
I&
$2A_{g}$&
2.86&
$|H\rightarrow L+1\rangle+c.c.\:(0.57)$\tabularnewline
&
&
&
&
 $|H\rightarrow L;H\rightarrow L\rangle(0.48)$\tabularnewline
\hline 
QCI&
II&
$4A_{g}$($mA_{g})$&
3.98&
$|H\rightarrow L+1\rangle+c.c.(0.37)$\tabularnewline
&
&
&
&
$|H\rightarrow L;H\rightarrow L\rangle(0.74)$\tabularnewline
\hline
QCI&
III&
$5A_{g}$&
4.44&
$|H-1\rightarrow L+2\rangle+c.c.(0.55)$\tabularnewline
&
&
&
&
$|H-2\rightarrow L;H\rightarrow L\rangle+c.c.(0.24)$\tabularnewline
&
&
&
&
$|H-1\rightarrow L;H\rightarrow L+1\rangle(0.23)$\tabularnewline
\hline
\end{tabular}\label{tab-pdpa5}
\end{table}

\begin{table}

\caption{Nature of $A_{g}$-type states contributing to the longitudinal TPA
spectrum of PDPA-10 in various CI calculations. Rest of the information
is same as given in the caption of table \ref{tab-pdpa5}.}

\begin{tabular}{|c|c|c|c|c|}
\hline 
Calculation&
Feature&
State&
Energy (eV)&
Wave Function\tabularnewline
\hline
\hline 
SCI&
I&
$2A_{g}$($mA_{g}$)&
2.32&
$|H\rightarrow L+1\rangle+c.c\:(0.68)$\tabularnewline
\hline 
QCI&
I&
$2A_{g}$&
2.08&
$|H\rightarrow L+1\rangle+c.c.\:(0.49)$\tabularnewline
&
&
&
&
 $|H\rightarrow L;H\rightarrow L\rangle(0.50)$\tabularnewline
\hline 
QCI&
II&
$3A_{g}$($mA_{g})$&
2.82&
$|H\rightarrow L+1\rangle+c.c.(0.68)$\tabularnewline
&
&
$4A_{g}(mA_{g})$&
2.88&
$|H\rightarrow L;H\rightarrow L\rangle(0.37)$\tabularnewline
\hline
QCI&
III&
$5A_{g}(mA_{g})$&
3.11&
$|H\rightarrow L+3\rangle+c.c.(0.42)$\tabularnewline
&
&
&
&
$|H\rightarrow L+1\rangle+c.c.(0.27)$\tabularnewline
&
&
&
&
$|H\rightarrow L;H\rightarrow L\rangle(0.33)$\tabularnewline
&
&
&
&
$|H-1\rightarrow L;H\rightarrow L+1\rangle(0.31)$\tabularnewline
&
&
&
&
$|H\rightarrow L+1;H\rightarrow L+3\rangle+c.c.(0.27)$\tabularnewline
\hline
\end{tabular}

\label{tab-pdpa10}
\end{table}

\begin{figure}
\begin{center}\includegraphics[%
  width=8cm,
  angle=-90]{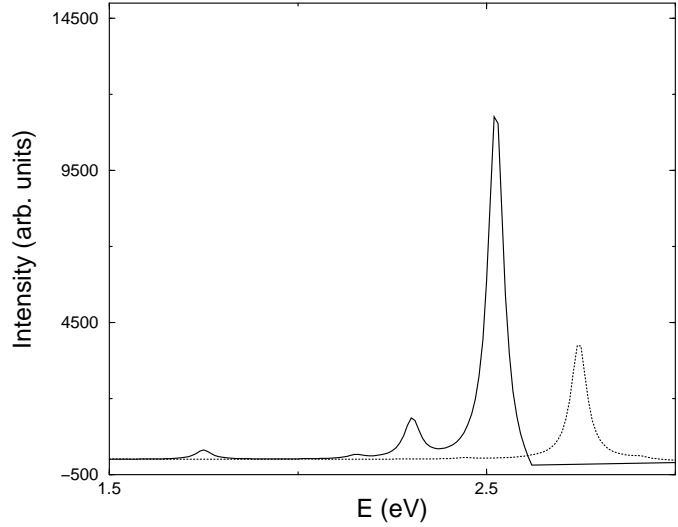}\end{center}

\caption{Comparison of the imaginary parts of $\chi_{xxxx}^{(3)}(-\omega;\omega,-\omega,\omega)$)
computed with the QCI method and the standard parameters for ten repeat
unit oligomers of: (a) PDPA (solid lines) and (b) \emph{trans}-polyacetylene
(dotted lines). A linewidth of 0.05 eV was assumed for all the levels.}

\label{tpa-pdpa-comp}
\end{figure}

Next, we compare the longitudinal component of the TPA spectra of
PDPA with that of \emph{trans}-polyacetylene for calculations performed
at the QCI level, using the standard parameters for the P-P-P Hamiltonian.
The imaginary parts of longitudinal TPA spectra of ten unit oligomers
of PDPA and \emph{trans}-polyacetylene are presented in Fig. \ref{tpa-pdpa-comp}.%
 The spectrum of \emph{trans}-polyacetylene oligomer has only one
significant peak corresponding to the $mA_{g}$ state while that of
PDPA-10 has again three visible features to which states $2A_{g}$,
$4A_{g}$, and $6A_{g}$ contribute. Although, the many-body wave
functions of these states are similar to the corresponding states
obtained with the screened parameters (see discussion above), however,
unlike the case for screened parameters, only one of the three peaks
($6A_{g})$ in the PDPA-10 spectrum is intense which clearly corresponds
to the $mA_{g}$, state of the oligomer. This is in excellent qualitative
agreement with the TPA spectrum of \emph{trans}-polyacetylene which
also shows only one intense peak corresponding to its $mA_{g}$ state.
As far as the quantitative comparisons between the two are concerned,
it is clear that the peak intensities in PDPA are significantly larger
as compared to the \emph{trans}-polyacetylene, and all the peaks in
the TPA spectrum of the PDPA are substantially redshifted as compared
to \emph{trans}-polyacetylene. It is noteworthy that we obtain this
result despite the fact that from the correlated calculations on PDPA-10
we have deleted all the orbitals except the ten most {}``chain-like''
orbitals. This clearly indicates that the presence of side phenyl
rings changes the nature of even the chain-like levels in such a way
that they lead to an enhanced nonlinear response. The results of these
correlated calculations also confirm the results of our H\"{u}ckel
model calculations presented in the previous section where we arrived
at the same conclusions regarding the longitudinal TPA spectrum of
PDPA vis-a-vis \emph{trans}-polyacetylene.

Thus the main conclusion of the present section is that as far as
the longitudinal TPA spectrum of PDPA is conerned, our calculations
suggest that, similar to the case of \emph{trans}-polyacetylene, it
will have one intense peak corresponding to the $mA_{g}$ state of
the system.

\subsubsection{Transverse Component}

\label{res-tpa-corr-y} Performing accurate correlated calculations
of $\chi_{yyyy}^{(3)}(-\omega;\omega,-\omega,\omega)$ for oligo-PDPA's
is an extremely difficult task. The reason behind this is that the
many-body $A_{g}$-type states which contribute to the peaks in this
component of the susceptibility are very high in energy, and thus
are more difficult to compute by many-body methods as compared to
the low-lying excited states contributing to the longitudinal spectra.
The very high excitation energies of these states are due to the fact
that these $A_{g}$-type states are predominantly composed of excited
configurations involving high-energy delocalized orbitals originating
from the side benzene rings. Therefore, it is clear that the SCI approach,
at best, may provide a qualitative picture of transverse nonlinear
optical properties of PDPA's, and, that too, for smaller oligomers.
For larger oligomers, results obtained with the SCI approach will
not be very reliable. However, during our independent-electron study,
we concluded that transverse susceptibilities exhibit rapid saturation
with the conjugation length. Therefore, we restrict our study of the
transverse components of the nonlinear susceptibilities to PDPA-5.
We perform calculations using the screened parameters both at the
SCI, as well as at the MRSDCI level. The SCI calculations were performed
using all the orbitals of the oligomer while, the more rigorous MRSDCI
calculations were performed using thirty orbitals in all, of which
fifteen were occupied orbitals, and the remaining fifteen were the
virtual ones. Rest of the occupied orbitals were frozen. Of the thirty
orbitals, ten were the orbitals closest to the Fermi level which were
also used in the QCI calculations. Remaining twenty orbitals were
the $d/d^{*}$-type phenylene-ring-based orbitals closest to the Fermi
level. In the MRSDCI calculations, no $l/l^{*}$-type phenylene orbitals
were used because they do not make any significant contribution to
the transverse nonlinear spectra either in the H\"{u}ckel model calculations,
or in the SCI calculations. In the MRSDCI calculations, we used 25
reference configurations for the $A_{g}$-type states, and 24 for
the $B_{u}$-type states leading to CI matrices of dimensions close
to half-a-million both for the $A_{g}$ and $B_{u}$ manifolds.

\begin{figure}
\begin{center}\includegraphics[%
  width=12cm,
  angle=-90]{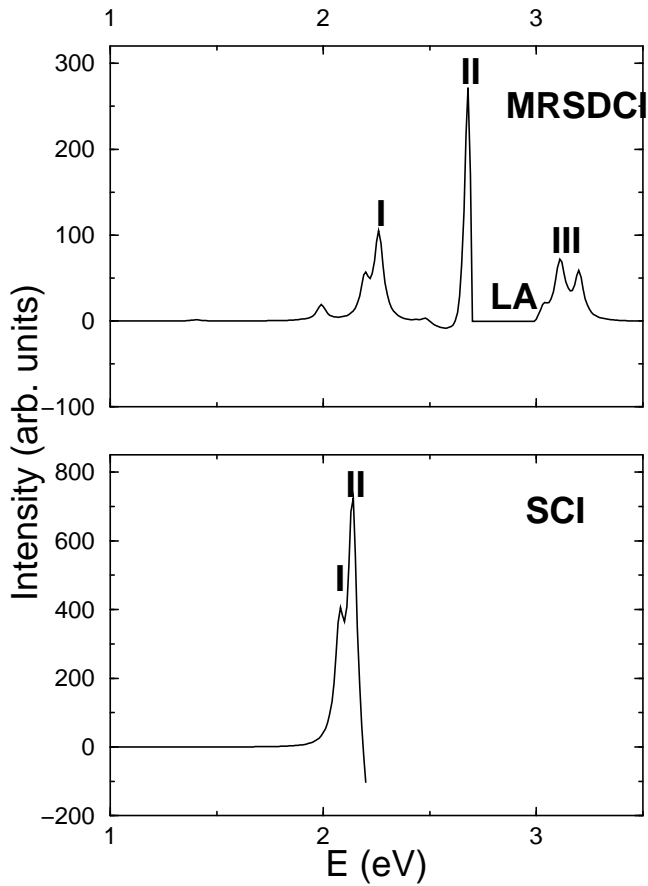}\end{center}

\caption{Imaginary part of $\chi_{yyyy}^{(3)}(-\omega;\omega,-\omega,\omega)$)
of PDPA-5 computed using the screened parameters and: (a) SCI method
(bottom), and (b) MRSDCI method (top) . A linewidth of 0.05 eV was
assumed for all the levels. }

\label{pdpa5_tpa_y}
\end{figure}

The transverse TPA spectra of PDPA-5 computed by us are presented
in Fig. \ref{pdpa5_tpa_y}.

The many-particle wave functions of various states contributing to
these spectra are presented in table \ref{tab-pdpa5-y}. The SCI spectrum
exhibits only one intense peak (feature II) accompanied by a shoulder
(feature I).%
 Upon examining the many-body states contributing to these peaks in
table \ref{tab-pdpa5-y}, we conclude that both these features are
due to $A_{g}$-type states whose wave function consists of single
excitations involving $L/H$ orbitals, and the high-energy $d/d^{*}$
class of orbitals ($H-17,H-19,L+17,L+19)$ localized on the benzene
rings. Thus, nature of states contributing to the transverse TPA spectrum
at the SCI level is the same as that obtained at the independent particle
level. 

Upon examining the MRSDCI spectrum of Fig. \ref{pdpa5_tpa_y}, we
see a significant redistribution of intensity over several peaks,
as compared to the SCI spectrum. As a result, the intensity of the
highest peak in the MRSDCI spectrum is much smaller as compared to
the highest one in the SCI spectrum. Next, we discuss the excited
states contributing to the three main features of the spectrum labeled
I, II, and III. The region labeled LA corresponds to the region of
the spectrum where $y$-polarized linear absorption dominates and
thus the actual TPA intensity there has large negative values. For
the sake of convenience, we have set the intensity in that region
to zero. Feature I, which is mainly due to the $6A_{g}$ state, is
composed primarily of single excitations involving $L/H$ and the
phenyl-based $d/d^{*}$ orbitals, and has a large coupling to the
$1B_{u}$ state via $y$-component of the dipole operator. Although
peak I of the MRSDCI spectrum is much weaker in intensity as compared
to its SCI counterpart, the nature of the many-body state contributing
to the peak in the MRSDCI spectrum is very similar to the ones contributing
to features I and II of the SCI spectrum.

\begin{table}

\caption{$A_{g}$-type states contributing to the transverse TPA spectrum
of PDPA-5 computed by various CI methods. Rest of the information
is same as given in the caption of table \ref{tab-pdpa5}.}

\begin{tabular}{|c|c|c|c|c|}
\hline 
Calculation&
Feature&
State&
Energy (eV)&
Wave Function\tabularnewline
\hline
\hline 
SCI&
I&
$9A_{g}$&
4.15&
$|H\rightarrow L+17\rangle+c.c\:(0.40)$\tabularnewline
&
&
&
&
$|H\rightarrow L+19\rangle+c.c\:(0.33)$\tabularnewline
\hline
SCI&
II&
$11A_{g}$&
4.28&
$|H\rightarrow L+17\rangle+c.c\:(0.38)$\tabularnewline
&
&
&
&
$|H\rightarrow L+19\rangle+c.c\:(0.54)$\tabularnewline
\hline 
MRSDCI&
I&
$6A_{g}$&
4.52&
$|H\rightarrow L+17\rangle+c.c\:(0.27)$\tabularnewline
&
&
&
&
$|H\rightarrow L+19\rangle+c.c\:(0.38)$\tabularnewline
&
&
&
&
$|H-2\rightarrow L+1\rangle+c.c\:(0.30)$\tabularnewline
\hline 
MRSDCI&
II&
$12A_{g}$&
5.36&
$|H-1\rightarrow L+2\rangle+c.c.(0.34)$\tabularnewline
&
&
&
&
$|H-1\rightarrow L+1;H\rightarrow L\rangle(0.39)$\tabularnewline
&
&
&
&
$|H-1\rightarrow L;H-1\rightarrow L\rangle+c.c.(0.35)$\tabularnewline
\hline
MRSDCI&
III&
$17A_{g}$&
6.25&
$|H\rightarrow L+19\rangle+c.c.(0.35)$\tabularnewline
&
&
&
&
$|H-1\rightarrow L+1;H\rightarrow L\rangle(0.39)$\tabularnewline
\hline
\end{tabular}

\label{tab-pdpa5-y}
\end{table}

Feature II is the most intense peak of the spectrum and it is due
to a high energy state $12A_{g}$ located at 5.36 eV which also has
a significant $y$-dipole coupling to the $1B_{u}$ state. However,
$12A_{g}$ state is qualitatively distinct from the $6A_{g}$ state
in that it exhibits strong mixing of singly- and doubly-excited configurations
involving low-lying orbitals, with no significant contribution from
the configurations involving the phenyl-based $d/d^{*}$ orbitals.
This result, which is qualitatively new when compared to the H\"uckel
model or the SCI results, is, clearly, an example of the superior
treatment of electron correlation effects by the MRSDCI approach,
because the double excitations are completely absent from the SCI
wave functions. Finally, feature III, which is also the last feature
of the MRSDCI spectrum, is mainly due the $17A_{g}$ state whose many-particle
wave function is also different from those of previous states in that
it is composed of double excitations involving low-lying orbitals,
and single excitations involving $H/L$ and phenyl-based $d/d^{*}$
orbitals.

\begin{figure}
\begin{center}\includegraphics[%
  width=12cm,
  angle=-90]{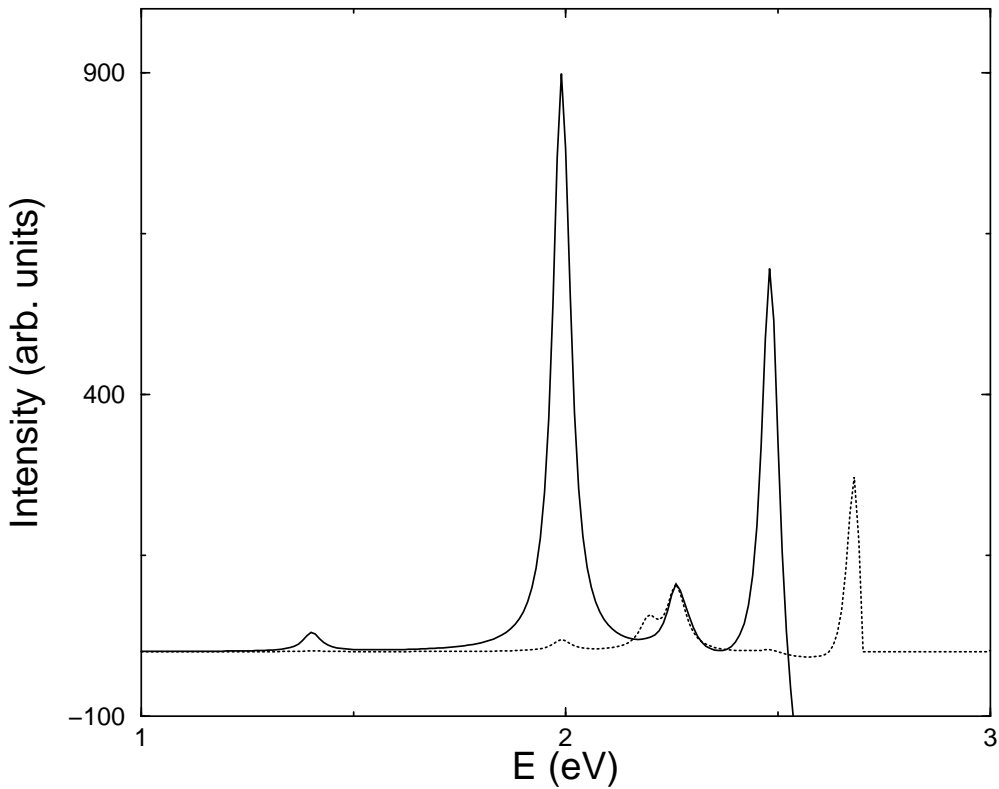}\end{center}

\caption{Comparison of longitudinal (solid lines ) and the transverse (dotted
lines) two photon absorption spectra of PDPA-5 computed using the
screened parameters and the MRSDCI method. A linewidth of 0.05 eV
was assumed for all the levels. }

\label{pdpa_tpa_x_y}
\end{figure}

In Fig. \ref{pdpa_tpa_x_y} we compare the MRSDCI transverse TPA spectrum
of PDPA-5 with its longitudinal one, and find that although transverse
TPA resonant intensities are considerable, but they are weaker than
their longitudinal counterparts.%
 Clearly, the height of the most intense peak in the transverse spectrum
is about one-fourth the height of the strongest peak of the longitudinal
spectrum. We speculate that the reason behind this result is that
electron correlation effects redistribute the intensity in the transverse
spectrum over several peaks, making the resonant response weaker.
Therefore, it will be of considerable interest to perform correlated
calculations on longer oligomers to check whether aforesaid results
hold true in the bulk limit as well.

\section{Conclusions and Future Directions}

\label{conclusion}

Our aim behind undertaking the present theoretical study of the nonlinear
optical properties of the novel polymer PDPA was, not only to calculate
its TPA spectra, but also to understand them in a way similar to what
has been possible for simpler polymers such as \emph{trans}-polyacetylene\cite{dixit,sumit},
PPP, and PPV\cite{lavren,hng-ppv}, i.e. in terms of an essential
state mechanism involving a small number of excited states. For PDPA's,
whose structure has ingredients in common with both the chain-like,
as well as the phenyl-based polymers, our calculations suggest that
their longitudinal nonlinear optical properties can certainly be understood
in terms of an essential state mechanism, with the $mA_{g}$ state
being the dominant feature in the spectrum. Thus, so far as the longitudinal
TPA spectrum of oligo-PDPA's is concerned, the system behaves as though
it were a polyene with different hopping and electron-repulsion parameters
so as to account for its lower optical gaps, and enhanced resonant
response. 

As far as the tranverse spectrum is concerned, our calculations performed
on PDPA-5 suggest that there are three main features contributing
to the TPA spectrum, if sophisticated correlated calculations are
performed. Therfore, it is of considerable theoretical interest to
examine whether for longer oligomers, several $A_{g}$-type states
will make important contributions to the transverse TPA spectrum,
or only one state (the transverse $mA_{g}$ state) will eventually
survive. The other interesting aspect is that whether the transverse
$mA_{g}$ state will owe its origins to the phenyl based orbitals
as per H\"uckel model and SCI calculations, or will it emerge from
the configurations interaction effects involving excitations among
low-lying orbitals, as appears to be the case with the MRSDCI results
on PDPA-5. These questions can only be answered by performing calculations
on longer oligomers of PDPA's which we intend to pursue in future.

\section{Acknowledgments}

These calculations were performed on the Alpha workstations of Physics
Department, and the Computer Center, IIT Bombay. We are grateful to
Sumit Mazumdar (University of Arizona) for a critical reading of the
manuscript, and for many suggestions for improvement.


\begin{thebibliography}{10}
\bibitem{chi-explan}Clearly, because of anisotropy, in these materials one can obtain
longitudinal nonlinear optical response due to transversely polarized
radiation also. However, such response, which will be possible due
to the off-diagonal components such as $\chi_{xxyy}^{(3)}$ etc.,
will be much smaller.
\bibitem{notation}The many-particle wave functions of half-filled conjugated polymers
such as PDPA exhibit particle-hole (charge conjugation) symmetry when
treated using the models such as the Hubbard model and the P-P-P model.
Therefore, for every configuration which contributes to its wave function,
its charge-conjugated (c.c.) counterpart will contribute in equal
measure. For example, if configuration $|H\rightarrow L+1\rangle$
occurs in the wave function with coefficient 0.69, configuration $|H-1\rightarrow L\rangle$
will also occur in the wave function with coefficient of the same
magnitude (although, the sign could be reversed). Thus, we will use
shorthand notation $|H\rightarrow L+1\rangle+c.c.$ (0.69) to denote
the contribution of these configurations to the wave function.
\end{thebibliography}
\end{document}